\documentclass[sigconf, table]{acmart}

\usepackage{booktabs} 
\usepackage{multirow} 
\usepackage{amsmath}
\usepackage{graphicx}
\usepackage{subcaption}
\usepackage{array}

\AtBeginDocument{%
  \providecommand\BibTeX{{%
    \normalfont B\kern-0.5em{\scshape i\kern-0.25em b}\kern-0.8em\TeX}}}

\setcopyright{acmlicensed}
\copyrightyear{2024}
\acmYear{2024}
\acmDOI{XXXXXXX.XXXXXXX}

\acmConference[Conference acronym 'XX]{Make sure to enter the correct
  conference title from your rights confirmation emai}{June 03--05,
  2018}{Woodstock, NY}
\acmBooktitle{Woodstock '18: ACM Symposium on Neural Gaze Detection,
 June 03--05, 2018, Woodstock, NY} 
\acmISBN{978-1-4503-XXXX-X/18/06}

\definecolor{1}{RGB}{127,201,127}
\definecolor{2}{RGB}{190,174,212}
\definecolor{3}{RGB}{253,192,134}
\definecolor{4}{RGB}{255,255,153}
\definecolor{5}{RGB}{56,108,176}
\definecolor{6}{RGB}{240,2,127}
\definecolor{7}{RGB}{191,91,23}
\definecolor{8}{RGB}{102,102,102}

\begin{document}

\title{Beyond Gaze Points: Augmenting Eye Movement with Brainwave Data for Multimodal User Authentication in Extended Reality}

\renewcommand{\shorttitle}{Beyond Gaze Points: Multimodal User Authentication}
\author{Matin Fallahi}
\affiliation{%
  \institution{KASTEL Security Research Labs, KIT}
  \city{Karlsruhe}
  \country{Germany}}
\email{matin.fallahi@kit.edu}

\author{Patricia Arias-Cabarcos}
\affiliation{%
  \institution{Paderborn University and KASTEL Security Research Labs}
  \city{Paderborn}
  \country{Germany}}
\email{pac@mail.upb.de}

\author{Thorsten Strufe}
\affiliation{%
 \institution{KASTEL Security Research Labs, KIT}
 \city{Karlsruhe}
 \state{Arunachal Pradesh}
 \country{Germany}}
\email{strufe@kit.edu}

\renewcommand{\shortauthors}{anonymous and anonymous, et al.}

\begin{abstract}
    Extended Reality (XR) technologies are becoming integral to daily life. However, password-based authentication in XR disrupts immersion due to poor usability, as entering credentials with XR controllers is cumbersome and error-prone. This leads users to choose weaker passwords, compromising security. To improve both usability and security, we introduce a multimodal biometric authentication system that combines eye movements and brainwave patterns using consumer-grade sensors that can be integrated into XR devices. Our prototype, developed and evaluated with 30 participants, achieves an Equal Error Rate (EER) of 0.29\%, outperforming eye movement (1.82\%) and brainwave (4.92\%) modalities alone, as well as state-of-the-art biometric alternatives (EERs between 2.5\% and 7\%). Furthermore, this system enables seamless authentication through visual stimuli without complex interaction.
\end{abstract}





\maketitle

\section{Introduction}

Extended reality\footnote{'Extended Reality' is also known as xReality in some papers. The term 'Extended Reality' was used due to its widespread usage in literature.} (XR) is a collective term that encompasses Virtual Reality (VR), Augmented Reality (AR), and Mixed Reality (MR), combining real and virtual environments for interactive user experiences \cite{ratcliffe2021extended,rauschnabel2022xr,fast2018testing}.
XR has emerged as a transformative technology, deeply embedding itself into various aspects of our daily lives—from education \cite{alnagrat2022review,elmqaddem2019augmented} and healthcare \cite{mathew2020role, makinen2022user} to entertainment \cite{clua2021workshop,ansari2022implementing} and social interaction \cite{mcveigh2021case,maples2017use}. The unique attributes offered by XR (e.g., immersive 3D environments and real-time interactivity) are fundamentally changing how users experience and interact with content. In the evolving XR landscape, conventional authentication methods like passwords undermine the immersive experience. This situation underscores the demand for authentication methods that, while maintaining security levels comparable to or better than passwords, seamlessly integrate into the user experience \cite{stephenson2022sok}. One highly promising and viable alternative is found in biometric authentication systems \cite{stephenson2022sok}.

Biometric authentication systems employ unique traits like behavioral or physiological patterns to identify individuals. However, the authentication methods widely known from smartphones and desktops (e.g., facial recognition, fingerprint scanning, keystroke logging) are not seamlessly compatible with XR, e.g., they would require the use of external hardware or disrupt the immersive user experience. Meanwhile, XR devices are equipped with multiple sensors, including outward-facing cameras for environment-tracking and inward-facing cameras for eye-tracking to enhance user experience. 

In this unique environment, eye movements — naturally integrated into user interaction in XR \cite{zhang2018continuous} — offer a promising avenue for authentication. The use of eye movements as behavioral biometrics has already been explored in various authentication applications \cite{lohr2022eye,liebers2020gaze,eberz201928,sluganovic2018analysis,lohr2018implementation}. Eye-tracking techniques are non-intrusive and thus allow for hands-free interaction within the XR environment to ensure a seamless user experience. However, the existing literature on eye movement authentication demonstrates that the method is unreliable at the low frame rates commonly found in consumer-grade devices \cite{sluganovic2018analysis,lohr2022eye}.

In this paper, to improve the reliability of monomodal authentication systems based on eye movements, we adopted a multimodal approach that augments eye movement by measuring brainwave patterns as well. Multimodality has generally been shown to significantly improve authentication accuracy \cite{abinaya2022multimodal,bugdol2014multimodal}, and brainwave patterns, in particular, are a biometric which is unique to each individual and consequently resistant to spoofing, difficult to duplicate, and hands-free \cite{gui2019survey}. Since brainwave capturing is naturally hands-free, it is as suited to the XR setting as eye-tracking; however, the brainwave modality suffers from similar drawbacks as eye movement. Brainwaves are sensitive to noise and artifacts, especially in consumer-grade devices where performance is unreliable \cite{arias2021inexpensive,arias2023performance}. Therefore, we hypothesize that the integration of eye movement and brainwave modalities will yield a secure, robust, and user-friendly authentication mechanism that is fully compatible with XR environments.

To provide empirical validation for our multimodal authentication approach in XR, we conducted a lab study with 30 participants, using consumer-grade equipment to record both brainwaves and eye movements (Sec \ref{Section:EDP}). Leveraging this synchronized dataset, we develop an authentication system employing twin neural networks based (Sec \ref{Section:AE}) on an interactive dot stimulus \cite{sluganovic2018analysis}. 
We implemented various feature fusion and score fusion strategies to investigate the most promising configurations. Additionally, since the pupil diameter is used as a feature by some papers \cite{eberz201928,zhang2018continuous}, but others omit it \cite{sluganovic2018analysis,lohr2022eye}; We investigate the effect of pupil diameter on authentication performance to illuminate its exact influence on the system. The decision to include pupil diameter in feature engineering may be influenced by the designer's choice or the unavailability of pupil diameter.
Our results demonstrate a notably low Equal Error Rate (EER): 0.686\% without pupil diameter and 0.298\% with pupil diameter using score fusion (Sec \ref{Section:result}). Before concluding (Sec \ref{Section:conclusion}), we discuss the technical feasibility of XR integration (Sec \ref{Section:technical}), compare our results with related work (Sec \ref{Section:related}), and also discuss the limitations of our work (Sec \ref{Section:limit}). Accordingly, we offer the following contributions:

\begin{itemize}
            
            \item \textbf{Innovation in Multimodal Authentication:} Our authentication system is the first to combine synchronized eye movement and brainwaves. Thus, we offer a novel multimodal authentication solution tailored for XR.
            
            \item \textbf{Substantial Improvement in Accuracy:} We provide experimental confirmation that a multimodal authentication is more effective than a monomodal authentication in the XR context. Our authentication via eye movement shows a notable 81\%-83\% reduction in Equal Error Rate (EER) when augmented with brainwave data.

           \item \textbf{Insight into Pupil Diameter: } We investigate the impact of pupil diameter on authentication, offering additional depth to the understanding of eye movement results. The findings serve as a guide for feature selection in the authentication system based on eye movement.

\end{itemize}

\section{Authentication in XR}
Identity verification, or authentication, performs a 1:1 comparison to confirm if a presented biometric matches the registered individual, answering the question, \textit{"Are you who you claim to be?"} It is mainly used to differentiate legitimate users from attacker.
Authentication methods in XR can be broadly classified into three primary categories: Knowledge-Based methods \cite{mathis2020rubikauth,duzgun2022sok}, Possession-Factors methods \cite{chan2015glass}, and Inherence-Factors methods (biometrics) \cite{eberz201928,li2016whose,boutros2020iris}. Knowledge-Based methods involve what the user knows, like passwords or PINs. Possession-Factors utilize physical objects or tokens that the user owns, such as mobile phones or hardware dongles. Inherence-Factors, or biometrics, exploit unique physiological or behavioral traits for authentication.

In XR settings, authentication based on knowledge and possession factors are subject to usability challenges. Knowledge-Based methods often involve credentials that are difficult to remember and also difficult to enter, while authentication based on possession factors requires the availability of a physical token, the corresponding port on the device, and could be subject to theft \cite{stephenson2022sok}. Biometric authentication eliminates the need to memorize passwords or to carry external tokens for access. Further, biometrics have become a viable option for authentication in XR settings because new sensors are continually being integrated into XR devices to enhance the immersive experience for users \cite{stephenson2022sok}. 
 
While biometrics offer significant advantages in terms of usability, there are notable performance and accuracy issues that need to be addressed. Multimodal biometric authentication presents a potential solution to these challenges. Recently, eye movement and brainwave-based authentication have gained significant attention due to their hands-free operation, the difficulty of spoofing, and application beyond authentication \cite{arias2023performance,sluganovic2018analysis}. Therefore, combining eye movement and brainwave modalities could potentially mitigate accuracy challenges while maintaining their inherent advantages. Nevertheless, a biometric system requires the trait used in authentication to be universal (i.e., present in all individuals) and permanent (i.e., stable over time). Furthermore, the trait for authentication must also satisfy a whole line of further requirements. The trait must be unique and quantifiable, and it must also be resistant to forgery. The biometric trait must be revocable by the user, who will need to consider, as well, that authentication by that trait is acceptable. The price for sensing the trait must be affordable, and the performance of the sensing must be reliable.

Studies have demonstrated the temporal stability of brainwave and eye movement-based biometric authentication methods. Maiorana~\cite{maiorana2021learning} conducted a year-long study, confirming the viability of brainwave authentication. Similarly, Lohr et al.\cite{lohr2022eye} established that eye movement based authentication remains effective over a period of three years. While static biometrics, such as fingerprints or iris scans, cannot be altered if compromised. Lin et al.~\cite{lin2018brain} provided empirical data suggesting that brainwave passwords can be modified by assigning a different task. In terms of usability, several studies have indicated a general willingness among people to adopt eye movement and brainwave authentication methods, although privacy concerns still remain to be addressed~\cite{fallahi2024usability,rose2023overcoming,brooks2013perceptions}. Overall, we acknowledge the need for further research to examine other aspects of biometric authentication, but the specific aim of our work is to achieve reliable performance using sensors compatible with XR devices.

The reliability and performance of eye movement and brainwave authentication systems depend heavily on the quality of the recorded data. For instance, Lohr et al. \cite{lohr2022eye} demonstrated that when the sampling rate of an eye-tracker decreases from 1000Hz to 31Hz, the EER increases from 3.66\% to 23.37\%. Moreover, in eye movement authentication, Sluganovic et al. observed that a reduction in sampling rate from 500Hz to 50Hz resulted in an 11\% increase in the Equal Error Rate (EER). Similarly, in brainwave authentication systems, Arias-Cabarcos et al. \cite{arias2023performance} compared the performance of two different datasets: a consumer-grade dataset with a 256Hz sampling rate and a medical-grade dataset with a 1024Hz sampling rate. Using the same machine learning pipeline for both datasets, they found EERs of 8.5\% for the consumer-grade dataset and 1.9\% for the medical-grade dataset. Therefore, to effectively utilize brainwaves or eye movement for authentication, it is essential to enhance the performance of authentication systems in consumer-grade devices.

\section{Experimental Design and Procedures}
\label{Section:EDP}
Our research aims to achieve robust and reliable authentication in XR, specifically focusing on achieving high performance with consumer-grade devices. Since biometric authentication using just single modalities like eye movement~\cite{sluganovic2018analysis,lohr2022eye} and brainwaves~\cite{arias2021inexpensive,arias2023performance} have been shown to be unreliable, we investigate the efficacy of combining eye movement and brainwave data. Therefore, we formulate our central research questions as follows: Can synchronized eye movement and brainwave data improve performance compared with a single modality? Which modality is more reliable? What fusion strategy yields the best results? And, what impact does the feature of pupil diameter have on outcomes? To answer these questions, we have designed a set of experiments that is described in detail in this section.

\begin{figure}[h] 
\centering 
 \includegraphics[width=0.35\textwidth]{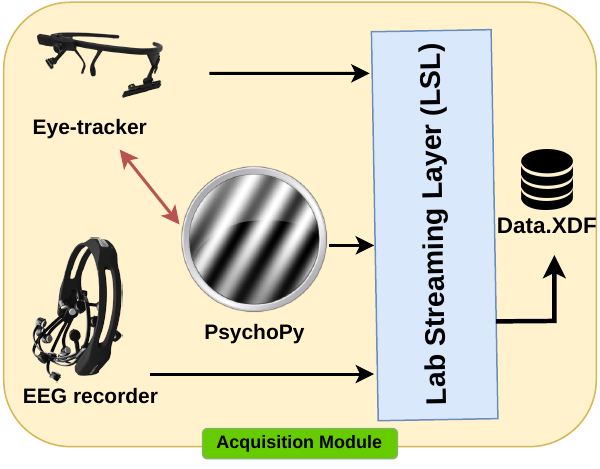} 
 \caption{\small Synchronized collection of eye-movment and brain activity. Lab Streaming Layer (LSL) receives data from PsychoPy, an eye-tracker, and a brainwave device. Also, PsychoPy and the eye-tracker coordinate to execute focus tasks, updating the display and sending focus timestamps to LSL} 
 \label{fig:lsl} 
\end{figure}

\subsection{Technological and Methodological Blueprint}
This section outlines the methodologies, software, and tools employed to implement multimodal authentication through the synchronized integration of eye movement and brainwaves in our experimental design (Figure \ref{fig:lsl}):

\begin{figure}[h] \centering \includegraphics[width=0.35\textwidth]{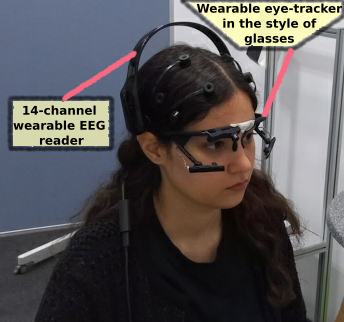} \caption{\small A participant in the experimental setup wearing the Pupil Core eye-tracker and the Emotiv EPOC X brainwave device} \label{fig:equipment} \end{figure}

\textbf{Authentication Task - Reflexive Saccadic Responses:} Building on  the methodology proposed by Sluganovic et al.\cite{sluganovic2018analysis}, we focused on measuring reflexive saccades due to their inherent stability and low susceptibility to temporary changes in mental or emotional conditions, such as attention, mood, or stress—referred to as transient cognitive states. Unlike voluntary saccades, reflexive saccades are driven by automatic mechanisms, making them more reliable for authentication. Our stimulus involved presenting a dot on the screen. When the participant's gaze fixated on the dot, it would disappear, and a new dot would appear in a random position. This sequence was repeated 25 times per round \cite{sluganovic2018analysis}, with participants completing 36 rounds. The number of rounds was determined based on the experiment's 25-minute time limit. To prevent fatigue and maintain focus, we included a 15-second rest interval between rounds. Previous studies by Sluganovic et al.\cite{sluganovic2018analysis} have demonstrated that this interactive dot task resists replay attacks effectively, as participants must respond to new random dot positions in real-time.

\textbf{Task implementation - PsychoPy:}  The authentication task was designed and executed using PsychoPy, a platform commonly used in neuroscience and psychology for creating complex visual and auditory stimuli \cite{peirce2009generating}. This choice was motivated by two key factors: first, its compatibility with our eye-tracker enabled the development of an interactive task that dynamically adjusts to the user's gaze position; second, its native Python support facilitated sending event markers to the software manage synchronized recording of our experiment.

\textbf{Synchronization - Lab Streaming Layer (LSL): }
Given the significance of millisecond-level precision in brainwave \cite{fernandez2017brain} and eye movement data \cite{belanger2015eye}, synchronization is crucial. We employed the Lab Streaming Layer (LSL)\footnote{https://github.com/sccn/labstreaminglayer}, which is already used in literature for synchronized multimodal recording of brainwave and eye movement data \cite{meier2018synchronized}. LSL is a system designed for unified time series data collection in research settings. We utilized LSL to achieve synchronized recording of brainwave and eye movement data, along with timestamps streams.

\textbf{Equipment:} In line with our objective to develop an authentication system for XR environments, we decided against using medical-grade EEG recorders or high-resolution desktop eye-trackers commonly employed in other research \cite{sluganovic2018analysis,eberz201928,fallahi2023brainnet,schons2018convolutional}. Instead, we selected an eyeglass-based eye-tracker and a neuroheadset tailored for general use. Our aim was to investigate the reliability of these devices for authentication tasks. The following section provides a technical overview of the equipment used (Figure \ref{fig:equipment}):

\begin{itemize}
    \item \textbf{Neuroheadset - Emotiv EPOC X:} The Emotiv EPOC X neuroheadset\footnote{https://www.emotiv.com/epoc-x/}, equipped with 14 EEG electrodes, records brainwave data at a sampling rate of 256 Hz. The accompanying  Emotiv software provides connectivity and quality metrics. The connectivity refers to how well the device connects to the head of the subject, while signal quality provides a summary measure that considers various factors such as movement, noise, signal amplitude, and other parameters. According to Emotiv's guidelines, optimal electrode contact with the scalp can yield a 100\% connectivity score. Nonetheless, achieving high-quality data can be challenging for individuals with long or thick hair.

    \item \textbf{Eye-Tracker - Pupil Core:} For a XR-like experience, we chose the Pupil Core\footnote{https://pupil-labs.com/products/core/}, as eye-tracker. Pupil Core includes one world camera, which records the surrounding environment and the participant's field of view, and two eye cameras that record the eyes of the participant simultaneously, capturing detailed information about gaze direction, pupil diameter, and eye movement, with its world cameras recording at 200 frames per second. To calibrate the eye-tracker, participants were asked to focus on a sequence of five dots on the screen using the Pupil Capture. The calibration process ensured the accuracy and alignment of the eye movement data with the participant's actual gaze direction.

\end{itemize}

\subsection{Participants and Ethical Considerations}
Our study included 30 participants, 11 women and 19 men. These participants were predominantly young adults (average age 24) affiliated with the university, either as undergraduate, master's, or PhD candidates or as research assistants. The only requirement for participation was that individuals be aged 18 or above.

The university's official social media channels were used to recruit such a cohort. We emphasized that participation was entirely voluntary, and participants had the freedom to withdraw from the study at any stage without any consequences. Participants were reimbursed 15 Euros per hour upon study completion as compensation for their time and contribution. 

This study adheres to responsible research practices by maintaining ethical integrity. All procedures, methodologies, consent form, and tools underwent rigorous scrutiny and were approved by our university's Institutional Review Board (IRB). Additionally, for Figure \ref{fig:equipment}, we obtained the subject's consent to use her photograph in the paper.

\subsection{Data Collection Process}
Before participating in an experiment, each subject must read and sign an informed consent form that explains the purposes of the experiment, the types of data to be collected, and how the data will be used. The preparatory step to the experiment itself is the equipment setup. The participating subjects first wear the Pupil Core eye-tracker recorder, and next, they wear the brain Emotiv EPOC X. The two devices are then adjusted for each subject, that is, the electrodes of the Emotiv EPOC X are placed to enhance data quality, and the Pupil Core eye-tracker is calibrated to achieve precise measurements of gaze points on the screen. To record data, we used the software Emotiv Pro to stream brainwave data and the software Pupil Player to stream eye movement data. Next, we launched the experiment script in PsychoPy to present stimuli and also to manage the event marking stream. Finally, we used the Lab Streaming Layer (LSL) to record data streams from the brainwave recorder, the eye-tracker, and PsychoPy (Figure \ref{fig:lsl}). Upon completion of the experiment, all devices are carefully removed from the participants, and compensation is provided in accordance with the informed consent agreement.

\begin{figure*}[ht]
\centering
   \makebox[\textwidth][c]{\includegraphics[width=0.9\textwidth]{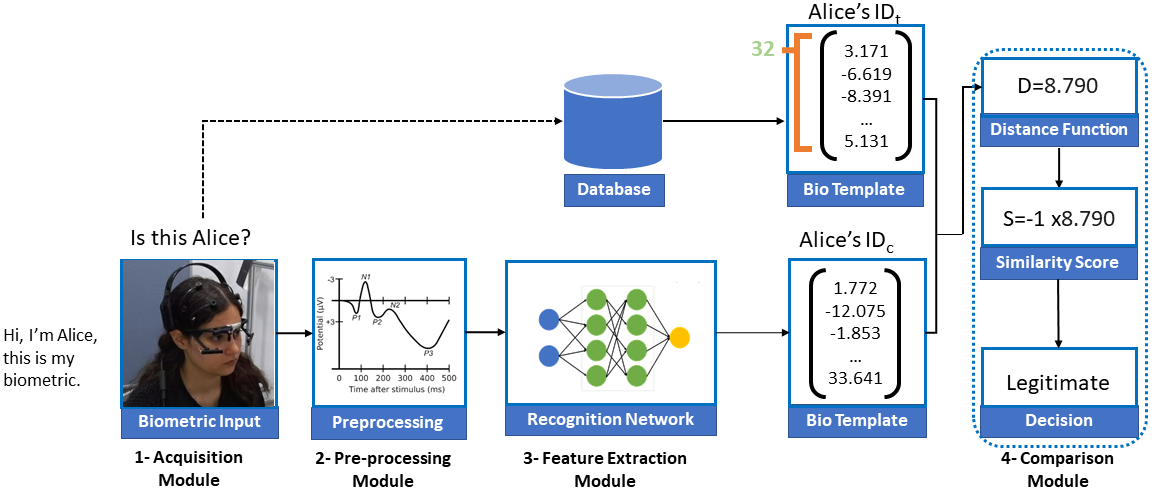}}%
  \caption{\small In the verification mode of our biometric recognition system, activity data from Alice is captured using a wearable device. This data undergoes pre-processing before entering the Feature Extraction Module, which condenses it into a 32-feature biometric template. Alice's claimed identity (IDc) is then compared with her pre-registered true identity (IDt), which consists of the biometric template established during the enrollment stage. Based on this comparison, the system's Comparison Module determines the legitimacy of Alice's identity.} 
  \label{fig:generalArchitecture}
\end{figure*}

\section{Authentication Architecture}
\label{Section:AE}
Here, we describe our authentication system architecture and fusion methods employed in the study.

\subsection{Authentication Approach Overview}
Figure ~\ref{fig:generalArchitecture} illustrates the verification model of a biometric system, which is based on four core building blocks: acquisition, preprocessing, recognition, and comparison. These components are universal and not tied to any specific biometric trait, serving as essential elements in any biometric authentication system.

\begin{enumerate}
    \item \textbf{Acquisition Module:} The initial step involves capturing the user's biometric data through specialized sensors. This phase is essential for collecting the raw information that will be analyzed and compared in subsequent stages. As detailed in the "Experimental Design and Procedures" (section \ref{Section:EDP}), the data acquisition phase involves the collection of user's data through specific sensors, tailored to the biometric trait being analyzed.
    
    \item \textbf{Pre-processing Module:} The acquired data undergoes a series of preprocessing steps to enhance its quality and make data ready for the next steps. This includes the extraction of relevant time series, filtering, interpolation, and standardization.
    
    \item \textbf{Recognition Module:} At this stage, the system identifies and extracts specific characteristics or features from the pre-processed data. These features are the unique attributes that distinguish one individual from another. The effectiveness of this step is paramount in ensuring that the system can accurately match the input data with stored templates. We used twin neural network, inspired by BrainNet \cite{fallahi2023brainnet}, to extract biometric features. The network's architecture, guided by a triplet loss function, effectively reduces the dimensionality of the data while preserving individual characteristics.
    
    \item \textbf{Comparison Module:} The extracted features are then compared to stored biometric templates. We used Euclidean distance (\( \sqrt{\sum_{i=1}^{n} (e_i - v_i)^2}\)) to compute the distance between enrollment and verification samples. Then it has been multiplied by -1, and the final similarity score is determined. Finally, a threshold should be set for deciding whether to accept or deny authentication requests.
\end{enumerate}

\subsection{Preprocessing}
\label{S4:pre}
The raw data includes multiple time series recorded during the experiment. In order to prepare input for our recognition module, we extract a segment of raw data known as a sample. This sample should consist of recorded biometric data when the subject is exposed to stimuli—in our case, a dot displayed on the screen. We extract these samples based on timestamps where the subject's gaze aligns with the dot's position on the screen.

\begin{enumerate}
\item \textbf{Data Extraction:} We extracted relevant time series data for brain, eye, and event timestamps from raw .xdf files, the output format of the Lab Streaming Layer (LSL). Specifically, for brain activity, we extracted 14 time series corresponding to 14 EEG electrodes. For eye movements, we extracted 12 time series related to the x and y coordinates of both the pupil and gaze point. We excluded time series related to the z-coordinate and gaze confidence, as they could introduce session-specific learning due to the fixed screen-to-user distance, which may vary based on the user's height and dependence of confidence based on environment factors and calibration. Also, we extracted 4 time series related to pupil diameter to investigate authentication with pupil diameter features.

\item \textbf{Sample Extraction:} Brainwave and eye-movement samples were extracted based on the timestamps corresponding to the last hit of the eye-gaze with the dots. We select a duration of 0.4 seconds for samples, which encompass 0.1 seconds before the event and extending 0.3 seconds after it. This duration is chosen to provide data between the dot hit moving to another dot. If a longer duration were selected, it would result in the inclusion of data from multiple dots within each sample, thereby complicating the analysis.

\item \textbf{Standardization:} We needed to have a fixed sample size as input for our recognition system. Therefore, through resampling, we achieved a consistent count of 256 data points per second across the dataset for both of eye-movement and brainwave data.

\item \textbf{Data Integrity and Reliability:} In the eye movement data, we had NaN values caused by blinking. We filtered samples abundant in NaN values to retain only high-quality samples. For the remaining eye movement samples containing NaN values, interpolation techniques were applied solely within the sample to prevent information leakage. Unlike eye data, the brain data didn't require this step, as the brainwave recorder software has a built-in interpolation mechanism.

\end{enumerate}

\subsection{Feature extraction: Twin Neural Network}
\label{ss:siames}
To authenticate subjects, we extract unique individual information from eye movement and brainwave signals, which often include noise. We employed a Twin Neural Network (TNN) with a triplet loss function as the core of our feature extraction module. A TNN is a specialized neural network architecture consisting of two or more identical sub-networks connected in parallel. The triplet loss function ensures that embeddings from the same identity are close, while embeddings from different identities are far apart ~\cite{schroff2015facenet}. The triplet loss function \( L \) is expressed as the Euclidean distance:

\begin{equation}
 \label{eqn:tripletlossfunction}
 L(A,P,N) = \max(\| (f(A)-f(P))\|^2 - \| (f(A)-f(N))\|^2 + \alpha, 0)
\end{equation}

Here, \( f \) denotes an embedding, \( A \) represents an anchor input, \( P \) is a positive input (sample for the same subject as \( A \)), and \( N \) is a negative input from a different subject. The parameter \( \alpha \) serves as a margin that enforces a minimum level of dissimilarity between positive and negative pairs, thereby enhancing the differentiation of samples. The objective is to minimize:

\begin{equation}
 \label{eqn:minimize}
 \sum_{n=i}^{N}\| (f(A_i)-f(P_i))\|^2 - \| (f(A_i)-f(N_i))\|^2 + \alpha
\end{equation}

The indices \( i \) correspond to the individual triplet inputs utilized during training. The selection of triplets adheres to a strategy inspired by FaceNet, promoting efficient convergence in learning~\cite{schroff2015facenet}.

We employed a CNN for our sub-network architecture due to its demonstrated effectiveness in various brainwave and eye movement authentication studies~\cite{fallahi2023brainnet,lohr2022eye, maiorana2021learning,bidgoly2022towards}, specifically using the CNN architecture proposed in the BrainNet paper\cite{fallahi2023brainnet}.

\subsection{Comparison in Verification Mode}
\label{ss:verification}
The next step is to make an authentication \textit{decision}. The decision-making process can either consider each pair of verification and enrollment samples independently or use a group of them collectively. The number of enrollment samples used can impact data acquisition and performance. Specifically, using fewer samples leads to quicker data acquisition times but compromises the accuracy of the system and vice versa. We explored various scenarios to gain deeper insights into these trade-offs.

\begin{itemize}
    \item \textbf{Fixed Threshold-One Sample (S1):} Fastest query time but potentially least reliable. For this scenario, We use one sample as verification and one sample as enrollment.
    
    \item \textbf{Fixed Threshold-Best Match (S2):}  In most cases, multiple enrollment samples are available for each subject, and the best match between the verification and all enrollment samples is chosen for decision-making. This aims to enhance the system's performance by leveraging multiple enrollment samples.

    \item \textbf{Tailored Threshold (S3): }Alternatively, setting an individualized threshold for each user may yield more accurate results tailored to the unique behavioral characteristics of each individual. This strategy is in line with many existing authentication systems that train a user-specific model~\cite{arias2023performance,nakanishi2019biometric, fallahi2023brainnet}. While this method is expected to achieve higher performance, it necessitates an initial calibration phase for each newly enrolled user to determine the optimal threshold in the real-world implementation.
\end{itemize}

\begin{figure}[h] 
\centering 
 \includegraphics[width=0.35\textwidth]{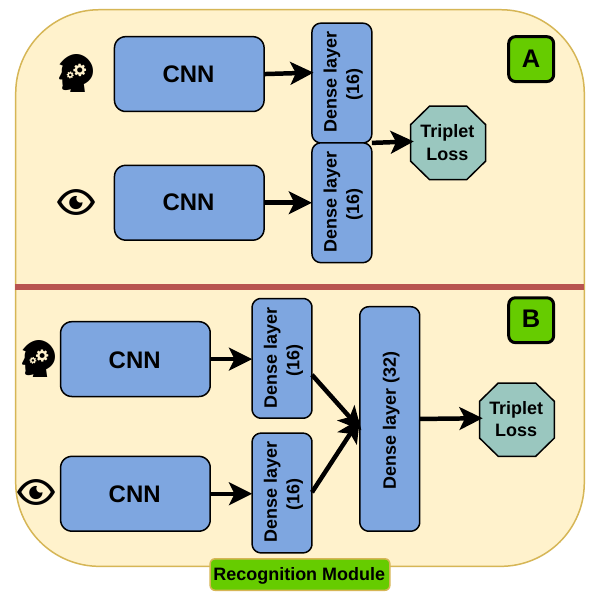} 
 \caption{\small Siamese sub-network architectures (A \& B) for eye movement and brainwave feature fusion.} 
 \label{fig:fusion} 
\end{figure}

\subsection{Fusion}
\label{ss:fusion}
In our biometric authentication system, fusion plays a pivotal role in integrating information from the two modalities we employ. We implement fusion at two different levels: at the feature and at the score levels. These fusion levels are designed to enhance the robustness and accuracy of the system by leveraging the complementary information present in both modalities. Below we detail the specific methods and considerations for each fusion level.

\textbf{Score Fusion:} To implement score fusion for multimodal biometric authentication, we trained separate twin neural networks for two distinct modalities: eye movement and brainwave data. These networks generate similarity scores that serve as the basis for subsequent fusion techniques. We use several established strategies to combine the similarity scores of eye movement and brainwave data pairs in the evaluation phase. Score fusion occurs in the \textit{Comparison Module} of the authentication system after similarity calculation for each modality. Specifically, the Max method computes the maximum score across each corresponding pair, symbolized as \( \max(s_{\text{{eye}}}, s_{\text{{brain}}}) \), where \( s_{\text{{eye}}}\) and \( s_{\text{{brain}}}\) represent the scores for eye and brain data, respectively. Conversely, the Min method calculates the minimum score using \( \min(s_{\text{{eye}}}, s_{\text{{brain}}}) \). The Average method takes the mean of both scores, expressed as \( \frac{{s_{\text{{eye}}} + s_{\text{{brain}}}}}{2} \). Finally, the product approach multiplies the scores, resulting in \( s_{\text{{eye}}} \cdot s_{\text{{brain}}} \). We employ these different fusion techniques to generate the final similarity score, thus making our authentication system more robust.

\textbf{Feature Fusion:} 
As depicted in Figure \ref{fig:fusion}, we employed two separate convolutional neural networks (CNNs) as subnetworks within a Siamese neural network for the purpose of feature fusion. In this way, we accommodate the different characteristics of brainwave and eye movement data. In \textit{Architecture A,} each of these CNN networks ends in a 16-dimensional dense layer. Subsequently, these two 16-dimensional layers are concatenated, resulting in a subject-representative layer comprising 32 values. On account of this structure, the feature fusion process is dynamically guided by the loss function, thus ensuring an integrated representation which accentuates the distinct attributes of both modalities. Moreover, in \textit{Architecture B}, to enhance the efficacy of the feature fusion process, we explored an alternative configuration that incorporates an additional 32-dimensional dense layer. This supplementary layer aims to facilitate a more complex integration of features, which, in its own right, will reduce total dependence on the loss function for effective fusion.

\begin{table*}
    \centering
    \begin{tabular}{clccc}
        \toprule
        \rowcolor[gray]{0.8}
        \multicolumn{2}{c}{\textbf{}} & \multicolumn{3}{c}{\textbf{Equal Error Rate}}\\
        \cmidrule(r){3-5} 
        \textbf{Approach} & \textbf{Biometric}& \textbf{S1} & \textbf{S2} &\cellcolor{6} \textbf{S3} \\
        \midrule
        \multirow{3}{*}{\textbf{Single Biometric}} 
        & Brainwaves                 & 14.55  & 5.560 & 4.920 \\ 
        & Eye-tracking with pupil    &\cellcolor{3}  13.42 &\cellcolor{3} 2.160 &\cellcolor{3} 1.820 \\
        & Eye-tracking without pupil & 19.07  & 3.732 & 3.639  \\
        \midrule[1pt]
        \multirow{5}{*}{\textbf{Score Fusion}} 
        
        & Max. with pupil & 13.00 & 4.355 & 3.763 \\
        
        & Min. with pupil & 13.35 & 1.481 & 1.231 \\
       
        & Mean with pupil & 8.574 & 0.507 & 0.385 \\
        
        & Product with pupil & \cellcolor{1} 7.098 & \cellcolor{1} 0.429 & \cellcolor{1} 0.298\\

        \cmidrule(l){2-5}
        
        & Max. without pupil    & 13.18   & 3.445 & 2.990\\
        
        & Min. without pupil    & 13.85   & 1.510 & 1.350\\
       
        & Mean without pupil    & 9.601   &\cellcolor{2} 0.850 &\cellcolor{2} 0.686 \\
        
        & Product without pupil &\cellcolor{2} 9.281   & 0.906 & 0.700 \\

        \midrule[1pt]
        \multirow{2}{*}{\textbf{Feature Fusion}} 

        & Architecture A with pupil &\cellcolor{4} 8.810 &\cellcolor{4} 0.917  &\cellcolor{4} 0.802\\
        
        & Architecture B with pupil & 13.36 & 1.831  & 1.674\\ 

        \cmidrule(l){2-5}
        
        & Architecture A without pupil &\cellcolor{7} 10.47 &\cellcolor{7} 1.550 &\cellcolor{7} 1.231\\
        
        & Architecture B without pupil & 16.28 & 4.373 & 4.085\\

        \bottomrule
    \end{tabular}
    \caption{This table displays the Equal Error Rate (EER) of our biometric authentication system based on three approaches: single modality, score fusion, and feature fusion. The columns represent different authentication strategies, specifically: S1 - Fixed Threshold with One Sample as the enrollment set; S2 - Fixed Threshold with the remainder of the samples used for enrollment; S3 - Tailored Threshold per subject with the remainder of the samples used for enrollment.}
    \label{T:EER}
\end{table*}
\section{Results and Testbed}
\label{Section:result}
This section delineates the experimental settings of the testbed and presents the outcomes across various metrics for different comparison strategies.

\subsection{Testbed and Evaluation Metrics}
\label{ss:testbed}

For a robust evaluation, we structured the dataset to segregate training and testing subjects. We employed 6-fold cross-validation, where each fold contained 25 subjects for training and an additional 5 unseen subjects for testing.  Also, during each fold of the cross-validation process, we learned a normalization function \footnote{https://scikit-learn.org/stable/modules/generated/
sklearn.preprocessing.StandardScaler.html} based on train data and normalized test and train data based on it.

The evaluation data were analyzed using the comparison scenarios detailed in Section \ref{ss:verification}.
To ensure the integrity of the evaluation, samples originating from the same experimental round as the verification samples were excluded. Consequently, enrollment and verification samples were consistently derived from two separate rounds of the experiment.

\textbf{Metrics:} The Equal Error Rate (EER) served as a summary metric, indicating the point where the False Acceptance Rate (FAR) and the False Rejection Rate (FRR) are equal.
Additionally, we report FRR at specific FAR thresholds of 1\%, 0.1\%, and 0.01\%.  The FAR represents the success rate of an attacker in a zero-effort attack, and the goal is to achieve a lower FAR while maintaining a reasonable FRR. Moreover, high FRR may lead to additional verification attempts, which can harm the device's usability. Therefore, a balance must be maintained between these two metrics to ensure robust security without compromising the user experience.  

The EER serves as a useful comparison metric across studies but is not directly practical for real-world applications. Notably, NIST (2023)/ISO~\footnote{\url{https://pages.nist.gov/800-63-3/sp800-63b.html}} and the European Border Guard Agency Frontex~\footnote{\url{https://www.frontex.europa.eu/assets/Publications/Research/Best_Practice_Technical_Guidelines_ABC.pdf}} specify that biometric systems must operate at FAR $\leq 0.1\%$. Meanwhile, FIDO~\footnote{\url{https://fidoalliance.org/specs/biometric/requirements/Biometrics-Requirements-v4.0.1-fd-20240522.pdf}} and the upcoming NIST (August 2024)/ISO standards~\footnote{\url{https://pages.nist.gov/800-63-4/sp800-63b.html}} recommend an even stricter FAR $\leq 0.01\%$. Additionally, FIDO and late NIST/ISO standards propose an FRR $\leq 5\%$, ensuring 19 successful logins out of 20 attempts for legitimate users.

\begin{table*}
    \centering
    \begin{tabular}{clccccccccc}
        \toprule
        \rowcolor[gray]{0.8}
        \multicolumn{2}{c}{\textbf{}} & 
        \multicolumn{3}{c}{\textbf{FRR at FAR = 1\%}} &
        \multicolumn{3}{c}{\textbf{FRR at FAR = 0.1\%}} & \multicolumn{3}{c}{\textbf{FRR at FAR = 0.01\%}} \\
        \cmidrule(r){3-5} \cmidrule(l){6-8} \cmidrule(l){9-11}

        \textbf{Approach} & \textbf{Biometric}& \textbf{S 1} & \textbf{S 2} & \textbf{S 3} & \textbf{S 1}& \textbf{S 2} & \textbf{S 3} & \textbf{S 1} & \textbf{S 2} & \textbf{S 3} \\
        \midrule
        \multirow{3}{*}{\textbf{Single Biometric}} 
        & Brainwaves                 &  59.47   & 28.86 & 21.45&  83.64  & 61.22 & 49.09& 94.43 & 86.92  & 67.99 \\ 
        & Eye-tracking with pupil    &  52.85   & 8.149 & 6.852&  74.89  & 23.25 & 16.44& 86.17 & 42.08  & 25.00 \\
        & Eye-tracking without pupil &  68.73  & 17.10 & 17.97&  88.06  & 44.86 & 38.81& 96.08 & 66.33  & 52.80 \\
        \midrule[1pt]
        \multirow{5}{*}{\textbf{Score Fusion}}

        & Mean with pupil    &  25.29  & 0.346  & 0.356&  45.15 & 2.096 & 1.611&  62.71 & 6.214 &3.789   \\
        & Product with pupil &  21.88  & 0.236  & 0.239&  42.96 & 1.794 & 1.226&  60.99 & 6.995 &3.794   \\

        \cmidrule(l){2-11}

        & Mean without pupil    &  35.39 & 0.990 & 0.873&  59.03 & 5.857 & 5.442&  76.60   & 19.04  &11.61  \\
        
        & Product without pupil &  35.27 &1.123  & 0.965&  59.45 & 6.850 & 5.956&  77.20    & 18.72  &12.36  \\

        \midrule[1pt]
        \multirow{2}{*}{\textbf{Feature Fusion}} 

        & Architecture A with pupil    & 26.29  & 1.066 & 1.101& 48.99 &  6.062 & 5.262& 68.25 &  20.69  & 14.69\\

        \cmidrule(l){2-11}
        
        & Architecture A without pupil & 41.16  & 3.549 & 3.008& 65.98 & 15.58 & 10.80& 81.94 & 27.10  & 20.74 \\

        \bottomrule
    \end{tabular}
    \caption{FRR (\%) at different FAR for Single Biometric, Score Fusion, and Feature Fusion Approaches under Different Comparison Scenarios.
} 
    \label{T:FRM}

\end{table*}

\subsection{Threat Model}
In alignment with the methodology of Zhang et al.~\cite{zhang2024safari}, we consider an adversary whose goal is to access sensitive personal information—such as user accounts, photos, or financial data—or to perform unauthorized actions like initiating payments or installing malware on a user's XR device. We assume that the adversary is knowledgeable about the authentication dot task and has physical access to the user's XR headset. Given these assumptions and the adversary's available techniques, we classify the following attacks:

\textbf{Blind attack:} The adversary has no prior knowledge of the legitimate user's eye movement and brainwave patterns. To execute the attack, the attacker wears the user's XR headset and attempts authentication with their own biometric samples. However, since the attacker cannot gain any advantage from observing the subject during authentication—due to brainwave data being completely resistant to observation and eye movement data requiring specialized devices—this attack is effectively equivalent to a mimic attack. In the mimic attack scenario, other threat models consider observers without any additional capabilities (unaided eye)~\cite{zhang2024safari}.

\textbf{Random input attack:}  We consider an adversary capable of circumventing the XR interface to gain access to the API of our biometric system, enabling them to input arbitrary feature vectors. The adversary's goal is to find a feature vector that is close to the genuine user's feature vector. Following Zhao et al.~\cite{zhao2020resilience}, we assume that the feature vectors are normalized and that the number of features is publicly known, with values between 0 and 1. To implement the attack, we generate 1 million samples with 32 values randomly selected from a uniform distribution between 0 and 1. Then, we compare them with the normalized feature vector of the legitimate users.

\subsection{Overall Results}
From our data collection, we obtained a total of 22,688 dot samples for analysis. Next, we trained the Siamese network (Sec \ref{ss:testbed}). The network was trained separately for brain and eye movement data. For the eye movement, we trained once considering pupil diameter and we trained again not considering pupil diameter. For both brain and eye data, we also applied two different feature fusion architectures (Sec \ref{ss:fusion}). The training and evaluation were conducted under the conditions specified in our testbed (\ref{ss:testbed}). The outcomes, particularly the Equal Error Rates (EER), are summarized in Table \ref{T:EER}.

\textbf{Fixed Threshold-Best Match vs. Tailored Threshold:} Our results clearly show that increasing the number of enrollment samples enhances the system's performance. However, the impact of different threshold strategies becomes evident when we focus on the scenarios which employ multiple enrollment samples, that is, the Fixed Threshold-Best Match scenario (S2) and the Tailored Threshold scenario (S3). For instance, using brainwaves, the EER drops from 5.560\% in the S2 to 4.920\% in the S3. This comparative analysis reveals that S3 consistently outperforms S2 under the same conditions involving multiple enrollment samples. Moreover, the advantages of using a Tailored Threshold strategy become increasingly apparent when fusion methods are considered; for example, in Score Fusion methods, such as the product with pupil diameter, the EER improves from 0.429\% in S2 to 0.298\% in S3 indicates the 30\% reduction in the Error.

\textbf{Single Biometrics vs. Fusion Approaches:} Table \ref{T:EER} reveals a consistent advantage for Fusion Approaches over Single Biometrics across various scenarios, particularly in the most advanced scenario (S3). For Single Biometrics, Eye-tracking with pupil diameter registers the lowest EER, achieving 1.82\%. However, this is significantly outperformed by Score Fusion methods, such as the product with pupil diameter, which exhibits an EER of just 0.298\%. Likewise, Feature Fusion's Architecture A with pupil demonstrates superior performance with an EER of 0.802\%. These results confirm that Fusion Approaches markedly outshine Single Biometrics.

\textbf{Score Fusion vs. Feature Fusion:}
Score fusion generally outperforms feature fusion across multiple comparison scenarios. Within score fusion, the mean and product strategies demonstrate a marked advantage over the min and max approaches in reducing the EER. On the feature fusion front, Architecture A consistently yields better results than Architecture B, although neither matches the high performance of score fusion techniques employing mean or product strategies.

\textbf{Influence of Pupil Diameter:}
The table \ref{T:EER} clearly illustrates the role of pupil diameter in enhancing authentication performance. When comparing eye-tracking methods with and without pupil diameter, there's a consistent improvement in EER across all examined scenarios. For instance, in the S3 scenario, eye-tracking with pupil diameter data yields an EER of 1.820\%, whereas the approach without pupil diameter data results in a higher EER of 3.639\%.

\textbf{False Acceptance Rate (FAR) Insights:}
In practical applications, EER is often not the main metric of focus. Instead, the practical usage needs to minimize FAR to bolster system security against zero-effort attacks while maintaining a reasonable FRR To elucidate the trade-offs between these metrics, we present data in Table \ref{T:FRM}. A comparison between FAR at 0.01\%, 0.1\%, and 1\% reveals that single-biometric approaches suffer from high FRRs, particularly in stringent security settings with low FARs (0.1\% and 0.01\%). For instance, in the S3 scenario with a FAR of 0.01\%, FRRs for single-biometric approaches like Brainwaves and Eye-tracking with pupil are 67\% and 25\%, respectively. These high FRRs indicate that single-biometric methods could be impractical for high-security applications. In contrast, multi-modal methods, especially those utilizing score fusion, significantly alleviate this issue. For example, the Mean Score Fusion with and without pupil diameter method results in a much lower FRR of 3.789\% and 11.61\% at a 0.01\% FAR, showcasing its effectiveness in balancing security and usability.

\begin{figure*}[h] \centering \includegraphics[width=0.80\textwidth]{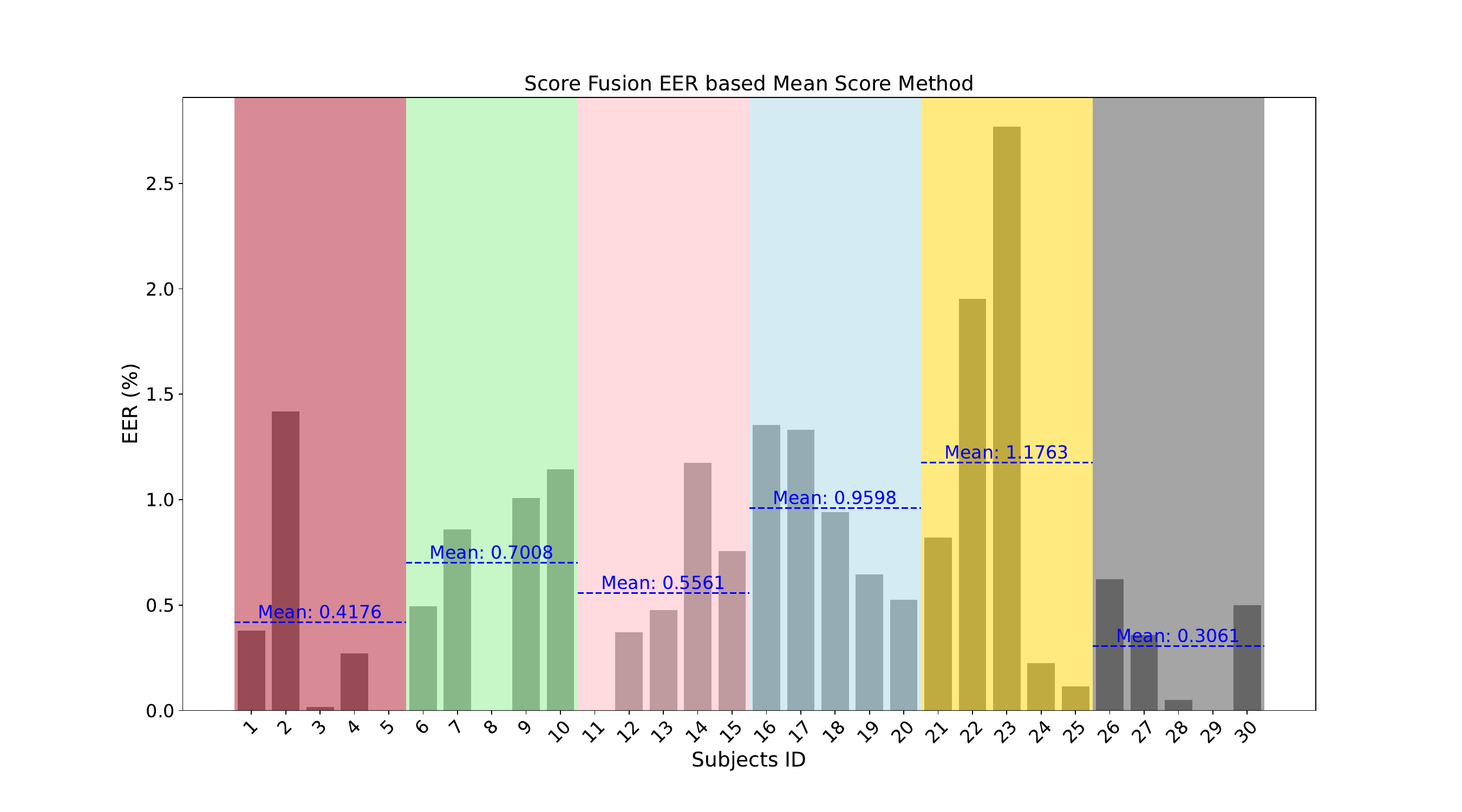} \caption{\small The bar plot illustrates the EER of test subjects, computed using the Mean Score Fusion without pupil method across multiple rounds of cross-validation, denoted by varied background shades. The dashed blue lines indicate the mean EER for each set of 5 subjects in the same cross-validation rounds.} \label{fig:S3eermean} \end{figure*}

\textbf{Insight into Subject-Level EER:} Figure \ref{fig:S3eermean} reveals distinct patterns in user-level EER, with values ranging from 0 to 2.77\% On average, the EER stands at 0.686, but a relatively high variance of 0.397 suggests notable differences in authentication performance across subjects. Specifically, about eight subjects exhibit EERs that are close to zero, underlining the system's effectiveness for these individuals. Conversely, five subjects manifest EERs nearly twice the average, accounting for the high variance and indicating that the system may require optimization for these cases. In terms of cross-validation rounds, some models appear to perform better than others, or certain subjects may have noise in their data samples. Interestingly, even within rounds that have a higher average EER, some subjects still achieve low EER values. This suggests that the higher error rates are likely not a result of model inefficiency but rather may stem from noise in the data samples for specific subjects.

\begin{figure}
    \centering
    \includegraphics[width=1\linewidth]{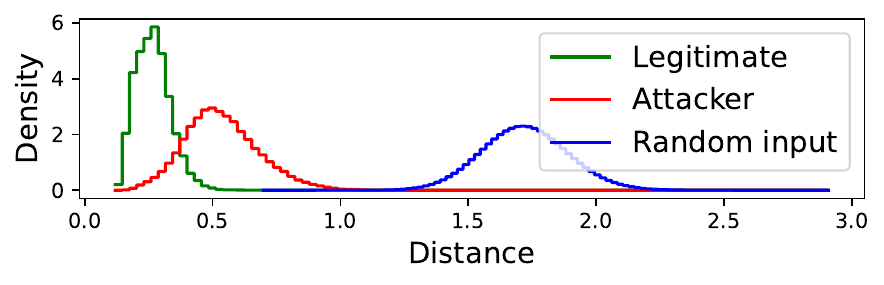}
    \caption{ Density distribution of brainwave similarity scores for legitimate users, human attackers, and random input features (it is similar for eye movement and fusion). }
    \label{fig:randomia}
\end{figure}

\textbf{Random input attack}
The density distance plot (Figure~\ref{fig:randomia}) demonstrates that the distances between randomly generated features and legitimate users are significantly greater than those between human attackers and legitimate users. This observation confirms that distance-based recognition systems are robust against random feature input attacks~\cite{pagnin2014leakage,zhao2020resilience}, suggesting that such attacks are less effective than blind attacker scenarios.

\textbf{Correlation between modalities}: to investigate the correlation between synchronized eye movements and brainwaves, we applied the Pearson correlation coefficient ~\cite{cohen2009pearson} for time domain analysis and magnitude-squared coherence, calculated using Welch's method~\cite{welch1967use}, for frequency domain analysis. We explored correlations between: (1) different channels of eye movement and brainwave data separately; (2) synchronized brainwave and eye movement data; (3) brainwave and eye movement data across different times of the experiment; and (4) brainwave and eye movement data from different subjects. The analysis revealed a consistent trend in both the time and frequency domains, demonstrating a strong\footnote{Higher than 0.6} correlation within the eye movement data and within the brainwave data (1). However, cross-modality correlations between eye movements and brainwaves were generally weak\footnote{Lower than 0.1}. Notably, the correlation of synchronized data from the same subject (2) was approximately 1.5 times stronger than that observed in cross-subject and time comparisons (3-4). The synchronized correlations demonstrated a more significant association between x-axis eye movement features and the frontal lobe region of the brain, particularly in the right hemisphere.

In summary, the weak correlation in synchronized data and strong correlation within modalities explain why the fusion of brainwave and eye movement data significantly improves the performance of our authentication system. Thus, we see our hypothesis confirmed that multimodal authentication based on eye movement and brainwaves substantially enhances the reliability of authentication using consumer-grade equipment with low sample recorder rates.

\section{Technical Feasibility of XR Integration and Usability Aspects}
\label{Section:technical}

To effectively integrate multimodal authentication into real-world applications, it is essential to consider both technical feasibility and usability. These two aspects are crucial for ensuring successful implementation and user acceptance. In the following, we discuss feasibility, based on the current and projected technological landascape; and usability, grounded on previous empirical studies evaluating similar interfaces.

\subsection{Technical Feasibility of XR Integration}
 Our multimodal authentication system can be readily adopted in XR environments because: (a) we deploy the system on separate but integrable devices, (b) we use a simple interactive dot task for implicit, hands-free authentication, and (c) we have selected two biometric modalities which are well suited for use in XR.

\textbf{Consumer-Grade Devices:} We aim to enhance the potential of consumer-grade devices for use in XR. We pursue this aim by employing only devices that are designed to operate at consumer-grade sampling rates.

For \textit{eye-tracking}, we used the Pupil Core device with a 200Hz sample rate in our experiments. Prior research by Pastel et al.~\cite{pastel2023application} demonstrates the viability of integrating such technology in XR,  citing 38 papers that used eye-trackers integrated with head-mounted displays (HMD). Moreover, companies like Pupil Labs and Tobii offer eye-tracking solutions designed for AR and VR devices~\footnote{https://www.tobii.com/products/integration/xr-headsets}. Additionally, Varjo provides a VR headset with an integrated eye-tracker operating at a 200Hz sample rate, aligning with our hardware specifications\footnote{https://varjo.com/products/vr-3/}.

Similarly, for \textit{brainwave recording}, we employed the Emotiv EPOC X, a consumer-grade device with a sampling rate of 256 Hz. Recently, Li et al.~\cite{li2023hair} demonstrated that EEG sponge electrodes can be seamlessly integrated into VR headsets. Additionally, dry EEG electrodes have been developed and are commercially available\footnote{https://www.bitbrain.com/neurotechnology-products/dry-eeg}\footnote{https://www.neurospec.com/Products/Details/1078/dsi-7} and some of these electrodes are already incorporated into VR devices\footnote{https://www.neurospec.com/Products/Details/1077/dsi-vr300}. Soon there will be VR headsets available that incorporate both EEG and eye-tracking capabilities\footnote{https://galea.co/}. 

\textbf{Dot Task:} To facilitate implicit authentication, we implemented an interactive dot task as the stimulus in our system. In a dot task, subjects are instructed to follow a dot displayed on the screen. This task offers high technical feasibility because it is simple and adaptable, thereby making it easy to integrate into daily routines or workflows. Consequently, our interactive dot task is well-suited for implicit authentication in the XR environment.

\textbf{Broad Applications:} We selected eye movement and brainwave data as our biometric modalities due to their extensive applications beyond mere authentication. These two modalities 
 are suitable for enhancing human-computer interactions in XR, as supported by existing studies \cite{gardony2020eye, aggarwal2022review}. The tracking of eye movements and brainwave patterns is also applicable in specialized entertainment contexts \cite{sundstedt2022gazing, de2023research}. Consequently, our chosen biometric modalities are particularly well-suited for integration into XR headset equipment.

\subsection{Usability Aspects of XR Integration}
Even with a fully implemented prototype; usability remains a pivotal factor for our proposed multimodal authentication system. To ensure the success of a biometric authentication system, it is crucial to understand user perspectives on the modality, its usability, and user concerns. Both brainwave and eye movement are emerging biometric modalities, and fully implemented solutions are not yet available in the market. Therefore, prioritizing usability is crucial as it greatly affects initial impressions and the subsequent adoption of these technologies.

In brainwave authentication usability studies, Chuang et al.\cite{chuang2013think} and Arias-Cabarcos et al.\cite{arias2023performance} found visual tasks more appealing than reading or auditory ones. Similarly, Röse et al.\cite{rose2023overcoming} and Fallahi et al.\cite{fallahi2024usability} confirmed a preference for visual tasks in their usability research with mockup prototypes. For eye movement, Brooks et al.\cite{brooks2013perceptions} showed that users found PIN entry simpler but considered the eye movement dot task more secure, generally preferring it over reading tasks and PIN entry. Fallahi et al.\cite{fallahi2024usability} also reported high usability scores of 78.8 in the SUS scale for the dot task, rated as "good" (A$^-$) based on Bangor et al.\cite{bangor2009determining} and Sauro et al.\cite{sauro11are}. Their results on eye movement and brainwave-based authentication indicate that users value usability, security, and passwordlessness as major advantages, while performance limitations and device overhead are seen as major disadvantages \cite{fallahi2024usability}. 

We selected the dot task as a visually appealing and usable task, and chose extended reality (XR) as the use case, where there is already an assumption of wearing a headset. The potential integration of brainwave and eye movement sensors into these headsets could eliminate the need for additional physical hardware for users. Thus, with a usable dot task and the elimination of extra hardware, our paper focuses on how the fusion of these two modalities could address performance concerns, which are a major consideration for users.

\begin{table*}[htb]\centering
\caption{\small Comparative analysis of eye movement authentication studies. }
\begin{tabular}{@{\extracolsep{2pt}}lccccc@{}}
\toprule
\footnotesize

 & \multicolumn{5}{c}{\textbf{Eye Movement Authentication}}\\                        

\multicolumn{1}{c}{Publication}& 
\multicolumn{1}{c}{Subjects Count}&
\multicolumn{1}{c}{Device} &
\multicolumn{1}{c}{Pupil Diameter} &
\multicolumn{1}{c}{Sample Rate} &
\multicolumn{1}{c}{EER}
\\

\cline{1-6}

				\textbf{Zhang et al. ~\cite{zhang2018continuous}, 2018}	&
                  30  & 
                  Glasses  & 
                 \checkmark &
				 50 Hz  & 
				 6.9\%  
				\\\hline
    
\textbf{Sluganovic et al. ~\cite{sluganovic2018analysis}, 2018}	&
30  &
Desktop  &
$\times$ &
500 Hz  &
6.3\% 

\\\hline

\textbf{Eberz et al. ~\cite{eberz201928}, 2019}	&
22  &
Desktop  &
\checkmark &
500 Hz  & 
1.88\% 
\\\hline 

\textbf{Lohr et al. ~\cite{lohr2022eye}, 2022}	&
322  &
Desktop  &
$\times$ &
1000 Hz  & 
3.66\% 
\\\hline

\textbf{Our work, Eye}	&
30  &
Glasses  &
$\times$ &
200 Hz  &  
3.64\% 	

\\\hline 

\textbf{Our work, Eye}	&
30  &
Glasses  &
\checkmark &
200 Hz  & 
1.82\% 	
\\\hline 

\textbf{Our work, Eye+EEG}	&
30  &
Glasses  &
\checkmark &
200 Hz  & 
0.298\% 	

\\\hline 

\bottomrule   

\end{tabular}

\label{tab:comparison}

\end{table*}

\section{Related Work}
\label{Section:related}
In the context of biometric authentication, our study distinguishes itself by focusing on consumer-grade devices that could used in XR settings. The following discussion elucidates how our contributions relate to existing work in three pivotal domains: brainwave authentication, eye movement authentication, and multimodal authentication approaches.

\subsection{Brainwave authentication} 
Considerable effort has been made to optimize performance in the domain of brain wave authentication. Nakanishi et al. \cite{nakanishi2019biometric} conducted an experiment with a sample size of 10 subjects and achieved a 4.4\% EER. Arias-Cabarcos et al.\cite{arias2021inexpensive} expanded the sample size to 50 subjects and achieved a 14.5\% EER. In a subsequent study, they improved their results to 8.5\% EER through enhanced machine learning techniques \cite{arias2023performance}. They used the same machine learning pipeline and a sample size of 40 subjects to achieve a 1.9\% EER on the medical dataset ERP CORE \cite{kappenman2021erp}. These advances in brain wave authentication demonstrate the significant impact of data quality on performance outcomes. Most recently, Fallahi et al.~\cite{fallahi2023brainnet} used again the ERP CORE dataset but with a triplet loss twin neural network to further improve performance to 1.37\% EER. Similarly, promising results are available on other medical datasets; for example, 0.14\% EER~\cite{fallahi2023brainnet}, 0.19\% EER~\cite{schons2018convolutional}, and 1.96\% EER~\cite{bidgoly2022towards}."

Our study utilizes the same network architecture as BrainNet~\cite{fallahi2023brainnet} to achieve a 4.92\% EER. Our performance surpasses that of Arias-Cabarcos et al.~\cite{arias2023performance}, and while it is essential to acknowledge that our EER is marginally higher by 0.5\% compared to the study by Nakanishi et al. \cite{nakanishi2019biometric}, we should note that, unlike our approach, Nakanishi et al. did not consider the unknown attacker \cite{mansfield2002best} scenario in their methodology. It is true that our performance did not reach the accuracy levels of medical-grade devices. However, our application is not in the medical sector but instead in XR, where integration of medical-grade devices, which are often bulky and complex to set up, is impractical.

\subsection{Eye movement authentication:}
In Table \ref{tab:comparison}, we present key parameters relevant to eye movement authentication in our related works, including device type (desktop or glasses), pupil diameter, number of subjects, EER, and sample rate.

\textbf{Pupil diameter as a feature:} Both Zhang et al. \cite{zhang2018continuous} and Eberz et al. \cite{eberz201928} incorporated pupil diameter as a feature into their models. In contrast, Sluganovic et al. \cite{sluganovic2018analysis} and Lohr et al.\cite{lohr2022eye} specifically excluded pupil diameter from their feature sets. Our research indicates that incorporating pupil diameter can enhance model performance by more than 50\%, thus making it an important factor to consider in comparisons and feature selection.

\textbf{EER Comparison:} Table \ref{tab:comparison} indicates that our EER outcomes closely align with the state-of-the-art results in single modality approaches. Our study yielded an EER of 1.82\% when pupil diameter data were incorporated and an EER of 3.64\% when not. Even despite the lowering sampling rate of our device, our results compare well to those presented by Eberz et al \cite{eberz201928} and Lohr et al. \cite{lohr2022eye}. We attribute this comparable performance to our refined comparison strategy. Specifically, our S3 strategy uses the remaining rounds for the enrollment set and adheres, as well, to a best-match scenario. This strategy effectively mitigates the effect of noise samples in the enrollment set. By contrast, when we use our simpler S1 strategy, the EER rates rise to 13.4\% and 19\%, which underscores the significant role played by sample rate in outcomes.

Further, our results also underscore the comparative advantage gained by incorporating brainwave data. Our EERs of 0.298\% and 0.686\% showcase the effectiveness of augmenting multimodal authentication, specifically by brainwave data.

\subsection{Multimodal authentication}
Multiple studies have been conducted on multimodal authentication systems; however, many focus on modalities that are not well-suited for XR environments. For example, Chakladar et al. \cite{chakladar2021multimodal} used EEG and signature-based methods, Zhang et al.\cite{zhang2020deepkey} combined EEG and gait, Wang et al. \cite{wang2016continuous} utilized EEG and face images, and Ammour et al. \cite{ammour2023multimodal} relied on Electrocardiogram (ECG) and fingerprints. In contrast, we found two studies more closely related to our work. First, Wu et al. \cite{wu2019lvid} explored the use of voice and lip movements as biometrics, achieving a 95\% True Positive Rate (TPR) and detecting 93.47\% of attacks (TNR) with 104 subjects. It is plausible to assume that XR devices could be equipped with a camera to capture lip movements. Second, Peng et al.\cite{peng2016continuous} proposed a system based on voice and hand motion, reporting a 99\% TPR and a 0.5\% FRR with 32 subjects.

For performance comparison, the referenced studies did not report EER, but we can make approximations based on FAR and FRR. In Wu et al.'s work~\cite{wu2019lvid}, they reported a 5\% FAR and 6.53\% FRR, leading us to conclude that their system has at least a 5\% EER. Similarly, Peng et al.~\cite{peng2016continuous} noted a 1\% FAR and 0.5\% FRR, suggesting a minimum EER of 0.5\% (probably in the middle of 0.5 and 1). Comparing these to our best results—0.686\% EER without pupil diameter data and 0.298\% EER with pupil diameter data—our system outperforms Wu et al. and is more effective than Peng et al. when considering the pupil diameter-based scenario. Moreover, unlike lip cameras, which serve no additional function, EEG and eye-tracking can be applied to various other applications, including human-computer interfaces \cite{aggarwal2022review} and entertainment \cite{de2023research}. When compared to voice and hand motion-based systems, we contend that our approach offers the advantage of being hands-free and potentially more effective in a crowded setting.

\textbf{In a somewhat related study}  Krishna et al.~\cite{krishna2019multimodal} explored the feasibility of using EEG and eye movement data for multimodal biometric authentication. Their approach involved combining two independent datasets related to eye movement and brainwave patterns to create a hypothetical multimodal dataset. While the idea is interesting, its practical applicability is doubtful. For the purposes of authentication, the point is not to amalgamate data from multiple subjects but to identify distinct characteristics which are unique to an individual. Furthermore, the results of Krishna et al.~\cite{krishna2019multimodal} showed poor performance in the eye movement modality, with a FAR of 7.4\% at an FRR of 36.7\%. These results led to no noticeable improvements over monomodal authentication. In fact, the only enhancement observed by the authors occurred in scenarios with 'low-confidence predictions of EEG.' Consequently, despite the innovative aspects of their approach, we do not categorize theirs as a multimodal authentication system based on eye movement and brainwave data.

\subsection{Comparison with VR Authentication Works}
\label{Section:related}
Reliable authentication in XR environment remains an open challenge. Biometrics have emerged as a viable solution; however, certain biometric methods require physical activities that may reduce practicality, ranging from discrete hand gestures~\cite{Peng2017-nh,Chauhan2014-ry} to more active ones like walking~\cite{Shen2019-je,Pfeuffer2019-ma} or throwing a virtual ball~\cite{Ajit2019-ao,Miller2019-yy}. Additionally, some solutions utilize sensors that may not be well-suited for XR environments~\cite{Chen2021-zz, Gao2019-of}; for instance, Chen et al.~\cite{Chen2021-zz} employed electrical muscle stimulation, which relies on sensors attached to the hands.

While several studies propose biometrics suitable for XR setups, such as free head and body movement~\cite{Mustafa2018-wo}, eye-related biometrics~\cite{zhang2018continuous, Luo2020-cx}, skull conductance~\cite{Schneegass2016-zc}, and brainwaves~\cite{Lin2018-dt}; their reported EERs range from 2.5\% to 7\%, comparable to our single-modality results (1.9\%-4.9\%). However, as shown in Table~\ref{T:FRM}
, higher security configurations result in increased FRR. Therefore, the multimodal approach can be a promising alternative to improve performance further.

Multi-factor approaches~\cite{Zhu2020-xy,lu2018multifactor}, such as combining biometrics with knowledge-based methods, aim to address performance challenges but inherit the limitations of both factors. An alternative is multimodal biometrics~\cite{peng2016continuous,Pfeuffer2019-ma}, which uses multi biometrics. We adopted this approach by fusing eye movement and brainwaves, reducing the error significantly.

\section{Limitations}
\label{Section:limit}
Our research faces two primary limitations: sample size and single-session data collection. First, although our sample size is comparable to that used in similar studies in this field \cite{eberz201928,sluganovic2018analysis,fallahi2023brainnet,wu2019lvid, zhang2018continuous}, it is relatively small when considering broader biometric research such as face and fingerprint recognition. Increasing the sample size could enhance our learning model and facilitate more realistic evaluations. Second, while relying on data from a single session is common in our research domain \cite{arias2023performance,nakanishi2019biometric,wu2019lvid,eberz201928,peng2016continuous}, this approach risks overfitting, which may degrade the performance of our methods in real-world applications over time. To address this, our experiment design avoids using samples from the same round for both enrollment and verification, and incorporates a 15-second rest period between rounds. While we anticipate a higher EER in multi-session scenarios \cite{seha2019new,lohr2022eye}, it is important to highlight that our primary contribution lies in demonstrating that the fusion of brainwave and eye movement can improve performance significantly, rather than achieving a specific error rate. Now that we showed feasibility, future work could explore robustness in bigger and varied datasets.

\section{Conclusion and Future Work}
\label{Section:conclusion}
In this study, our investigation shows that this combination of brainwaves and eye movement yields highly promising results. 
Through our research, we substantially improved authentication accuracy and enhanced resistance against zero-effort attacks. Specifically, our multimodal authentication system achieved an EER of 0.298\% and 0.686\%, along with FRR of 3.8\% and 11.6\% at FAR of 0.01\% compared with 25\% and 52.8\% FRR in single modality eye movement authentication. These results offer a higher level of security with a reasonable FRR, ensuring a smooth user experience without unnecessary disruptions. We provide a straightforward, hands-free authentication method that is both suitable for XR settings and appropriate for consumer-grade devices. Our multi-modal authentication system improves authentication accuracy while also holding promise for broader adoption in real-world applications.

In the future, investigating multi-session scenarios and increasing the sample size are essential steps for improving model learning and conducting more comprehensive evaluations. Furthermore, exploring additional tasks alongside the interactive dot task will contribute to a more extensive understanding of multimodal biometric authentication. Moreover, While the dot task can effectively resist against replay attacks, it would be beneficial to explore whether and how it is possible to use correlation in synchronized data to ensure that brainwave and eye movement data are recorded simultaneously, adding an extra layer of security.

\section*{Acknowledgments}
This work was funded by the Topic Engineering Secure Systems of the Helmholtz Association (HGF) and supported by KASTEL Security Research Labs, Karlsruhe. This work was also supported by Germany’s Excellence Strategy (EXC 2050/1 ‘CeTI’; ID 390696704). We thank the textician of KASTEL Security Research Labs for assistance and support in the research communication

\bibliographystyle{ACM-Reference-Format}
\bibliography{reference}

\end{document}